\documentclass[useAMS,usenatbib]{mn2e}
\usepackage{times}

\def\msun{\mbox{M$_{\odot}$}}

\def\spose#1{\hbox to 0pt{#1\hss}}
\newcommand\lsim{\mathrel{\spose{\lower 3pt\hbox{$\mathchar"218$}}
     \raise 2.0pt\hbox{$\mathchar"13C$}}}
\newcommand\gsim{\mathrel{\spose{\lower 3pt\hbox{$\mathchar"218$}}
     \raise 2.0pt\hbox{$\mathchar"13E$}}}

\begin{document}

\title[Search for early structure-formation signatures in HD molecule]{HD molecule 
and search for early structure-formation signatures in the Universe}
\author[N\'u\~nez-L\'opez, Lipovka \& Avila-Reese]{Ramona N\'u\~nez-L\'opez$%
^{1,2}$ \thanks{%
E-mail: ramona@astroscu.unam.mx}, Anton Lipovka$^{2}$ and Vladimir
Avila-Reese$^{1}$ \\
$^{1}$ Instituto de Astronom\'{\i}a, Universidad Nacional Aut\'onoma de
M\'exico, A.P. 70-264, 04510, M\'exico, D.F.\\
$^{2}$Centro de Investigaci\'on en F\'{\i}sica, UNISON, Rosales y Blvd.
Transversal, Col. Centro, Edif. 3-I, Hermosillo, Sonora, M\'exico, 83000}
\maketitle

\begin{abstract}
Possible detection of signatures of structure formation at the end of the
'dark age' epoch ($z\sim 40-20$) is examined. We discuss the 
spectral--spatial fluctuations in the CMBR temperature produced by 
elastic resonant scattering of CMBR photons on HD molecules 
located in protostructures moving with peculiar velocity. Detailed 
chemical kinematic evolution of HD molecules in the expanding homogeneous medium
is calculated. Then, the HD abundances are linked to protostructures at their 
maximum expansion, whose properties are estimated by using the top--hat spherical 
approach and the $\Lambda$CDM cosmology. We find that the optical depths in 
the HD three lowest pure rotational lines for high--peak protohaloes at their 
maximum expansion are much higher than those in LiH molecule. The corresponding 
spectral--spatial fluctuation amplitudes however are probably too weak as to be 
detected by current and forthcoming millimeter--telescope facilities.  
We extend our estimates of spectral--spatial fluctuations to gas clouds 
inside collapsed CDM haloes by using results from a crude model of HD production 
in these clouds. The fluctuations for the highest--peak CDM haloes at redshifts 
$\sim 20-30$ could be detected in the future. Observations will be important 
to test model predictions of early structure formation in the universe. 
\end{abstract}


\section{Introduction}

In the last years a great deal of interest arose for understanding the
formation of the first structures in the universe at the end of the
so--called 'dark age' epoch. This interest is motivated not only by the
possibilities of direct measurements of the physical conditions prevailing
in these (proto)structures and the constraining of cosmological parameters
at very high redshifts, but also by the measurement of the primordial
abundances of key elements (e.g., D and Li) in pregalactic epochs as a
direct signature of the Big Bang Nucleosynthesis (BBN). From the observational
side, the re--emission or absorption of Cosmic Microwave Background Radiation
(CMBR) photons in resonant lines from H$_{2}$, LiH, and HD, among other
primordial molecules, have been suggested as a viable way to detect cosmic
protostructures (Dubrovich 1977,1983; Maoli, Melchiorri \& Tosti 1994; 
Maoli et al.1996). Resonant absorption by neutral H at its 21 cm transition 
has also been proposed to trace protostructures at early cosmic times (Hogan 
\& Rees 1979; for more recent studies see Barkana \& Loeb 2005 and more 
references therein).

The first attempt to detect LiH emission and Doppler--induced anisotropies in
the CMBR from the protostructures at redshift $z=100$ was made with the 30
meter IRAM radio--telescope (de Bernardis et al. 1993). Only upper limits for
the abundance of LiH were obtained. It was believed that the relative
abundance of LiH, [LiH/H], at $z=30-100$ could be rather adequate (up to $%
10^{-10}$) to produce detectable spectral features due to the optical depth
in the lines of the LiH rotational structure. However, more detailed
calculations of the primordial LiH relative abundance have shown that, at
the discussed epochs ($z=30-100$), it should have been actually very small, 
[LiH/H] $\lsim10^{-18}$ (Stancil, Lepp \& Dalgarno 1996; Bougleux \& Galli 1997).
Therefore the optical depth in the rotational lines of LiH in the
primordial gas is too small to produce observational features (Galli \&
Palla 1998; Puy \& Signore 2001; Maoli et al. 2004).

The other primordial molecule, potentially interesting to detect features of
precollapse structures from the dark epoch, is deuterated hydrogen, HD. The 
dipole moment of the HD molecule is actually much smaller than the one of LiH: 
$d_{\rm HD}=8.3\times 10^{-4}$ debyes (Abgrall et al. 1982) and $d_{\rm LiH}=
5.88$ debyes. However, for [LiH/H]$\approx 10^{-18}$ and
[HD/H]=$(1-5)\ 10^{-9}$, one obtains in a first approximation
$\tau _{\rm HD}/\tau _{\rm LiH}\approx $ [HD]d$_{\rm HD}^{2}/$
[LiH] d$_{\rm {LiH}}^{2}\approx 20-100$, concluding that HD is a species 
more suitable for detecting the primordial structure signatures than LiH. 
Besides, the HD molecule is also able to amplify its emission due to
ro--vibrational luminescence effects (see e.g., Dubrovich 1997). At the
moment, the most remote of the detected absorbers in rotational lines of HD
molecule is at redshift $z=2.3377$ in the spectrum of the quasar $%
PKS1232+082 $ (Varshalovich et al. 2001). Nevertheless, the emission of this
molecule should be observable even from the pregalactic epoch.

In the intermediate epochs at $6\lsim z\lsim 30$, where nonlinear collapse takes
place for the mass scales of interest, the HD molecule can reemit the kinetic 
energy of gas out of equilibrium. The emission from primordial molecular 
clouds in the pure rotational structure of HD molecule owing to cooling 
processes has been calculated first time by Shchekinov (1986) and more accurate 
by  Kamaya \& Silk (2003). 
Recently, Lipovka, N\'{u}\~{n}ez--L\'{o}pez \& Avila--Reese (2005) have revisited 
the cooling function of this molecule  by taking into account its ro--vibrational
structure. At temperatures above $\sim 3000$K and at high densities, the 
cooling function is 1---2 orders of magnitude higher than reported 
before e.g., (Flower et al. 2000). If photo--ionization or shocks heat the 
dense collapsing clouds to 
temperatures above $\sim 1000$K, these new calculations show that the gas will 
be able to cool efficiently again with further HD resonant line emission,
and the intensity of the collapse energy reemission by the HD molecule calculated 
by Kamaya \& Silk (2003) will increase probably due to the ro-vibrational 
structure of this molecule. 
Recently, it was also suggested that the H$_2$ and HD fractions
can be substantially large in relic HII regions (Johnson \& Bromm 2006; 
Yoshida 2006), as well as in large mass haloes in which the gas is collisionally
ionized (Oh \& Haiman 2003).

At the linear stage of evolution of the first structures of interest 
($z>20-30$), the elastic resonant scattering of the CMBR by primordial
molecules in the evolving protostructures can damp CMBR primary anisotropies 
or, if the protostructures have peculiar velocities,
secondary anisotropies in emission (spectral--spatial fluctuations) can be
produced (Dubrovich 1977,1993,1997). In these cases, internal energy sources are
not required to produce the observational spectral and spatial features.
Therefore, {\it primordial resonant lines are ideal to study the evolution of
cosmic structures during their linear phase or the first phases of their
collapse, before star formation triggers}. The main goal of the present work
is to study the possible observational signatures from these phases previous
to the collapse of luminous objects in the Universe.

The HD molecule have at least two advantages for this task: 1) It is formed
rather fast in the homogeneous primordial gas at $z\approx 150$, through
channels similar to those of molecular hydrogen, and it may reach relatively
high abundances at those epochs. Indeed, at $z=100$, the relative abundance
of HD is already $\sim 10\%$ of the presend--day one. 2) The redshifted 
wavelength range of the rotational structure transitions of the HD molecule 
is very comfortable for detection. The first, second, and third rotational
transitions at the wavelength of 112.1 $\mu m$, 56.2 $\mu m$, and 37.7$\mu m$
can be detected at 0.8--3.5 mm if the lines are produced at redshifts around
20--30, typical of the epochs when the first stars probably started to form 
according to numerical and semi--analytical results (for a review see Bromm 
\& Larson 2004). Several
facilities planned or under construction will cover the submillimetric and
millimetric wavelength ranges with high sensitivity, for example the Large
Millimetric Telescope (LMT, or GTM for its acronym in Spanish) in Mexico,
whose radiometers cover the range from 0.8 to 3 mm.

Galli \& Palla (2002) presented results from detailed calculations of
chemical kinetics in thermally evolving primordial clouds, though
they did not follow the hydrodynamical evolution. The
relative abundance of the HD molecule that these authors found is very high
in the cooled gas clouds and, as it will be shown later here, must be
observable, so, more accurate calculations of the protocloud and cloud 
formation process and emission by the HD molecule are required.

The aim of this paper is twofold: to recall the importance of the HD molecule 
for investigating the very high--redshift universe, and to calculate the
parameters of HD lines of interest. The latter is a crucial step for
planning future observational strategies. We will estimate the optical
depth for protoclouds in the rotational lines of the primordial HD molecule
based on the $\Lambda$ Cold Dark Matter ($\Lambda$CDM) scenario of cosmic 
structure formation. For alternative scenarios (e.g., Khlopov \& Rubin 2004), 
the predictions will probably be different.

With the aim to obtain useful estimates, here we calculate detailed
chemical kinematics evolution for HD molecule in the expanding homogeneous 
medium (\S 2). Then we calculate the properties of high--density $\Lambda$CDM mass
perturbations (protohaloes) at their maximum expansion (\S 3.1), and estimate
the optical depths in the three lowest rotational line transitions of the HD
molecule fraction in these protohaloes (\S 3.2). We bear in mind that these
opacities will slightly decrease in the first phases of collapse when the
line--widths increase and become potentially observable as secondary spatial --
frequency anisotropies in the CMBR. We estimate the amplitudes and angular
sizes of these anisotropies assuming the linear model for the peculiar
velocities of the mass perturbations, and explore the possibility to detect
them by the LMT/GTM submillimeter telescope under construction in Mexico (\S %
3.3). Finally, we present a discussion of the implications of our results (%
\S 4).


\section{Evolution of the HD molecule abundance}


Following, we calculate the evolution of the HD abundance in the homogeneous
pregalactic medium. This is a good approximation to the HD abundance in
mass perturbations at their linear evolution phases as yet. We use the concordance 
cosmological model ($\Lambda$CDM) with $\Omega_{b,0}=0.04$, 
$\Omega _{M,0}=0.27$, $\Omega _{\Lambda }=0.73$ and $H_{0}=71$ kms$^{-1}$Mpc$^{-1}$%
(e.g., Spergel et al. 2003). Standard BBN yields for the
light elements are assumed.

In the early epochs, before the first cosmic objects are formed, the
thermodynamics of the primordial gas is out of equilibrium. Therefore, in
order to describe the molecular dynamics, we need to solve the kinetic
equation system, which can be written as follows: 
\begin{eqnarray}
\frac{dn_{i}}{dt}=\sum_{j,k}n_{j}n_{k}R_{jki}-n_{i}\sum_{m,n}n_{m}R_{imn}+ %
\cr \sum_{j}n_{j}R_{ji}-n_{i}\sum_{m}R_{im},  \label{kinetic}
\end{eqnarray}%
where $t$ is the cosmic time, $n_{i}$ is the density of the species $i$, $%
R_{jki}(T_{k})$ are the rates of the collisional processes $j+k\rightarrow i$
, as functions of the kinetic temperature $T_{k}$, $R_{ji}(T_{r})$ are the
rates of the radiative processes (formation and destruction of the molecule
by the CMBR photons characterized by the radiative temperature $T_{r}$).

In eq. (\ref{kinetic}) the positive and negative terms correspond to the
formation and destruction processes of the species $n_{i}$, respectively. To
solve the equations system (\ref{kinetic}), it is convenient to introduce
new variables: $x_{i}=n_{i}/n_{B}$, where $n_{B}=n_{B,0}\left( 1+z\right)
^{3}$ and $n_{B,0}$ is the present-day baryon number density, $n_{B,0}=\Omega
_{B}n_{cr}$. It is also convenient to substitute the cosmic time variable
with the redshift, by using the appropriate relation for the flat cosmology
with cosmological constant adopted here: 
\begin{equation}
dt=-\frac{1}{H_0}\frac{dz}{\left( 1+z\right)\sqrt{\Omega_{M,0}(1+z)^3 +
\Omega_{\Lambda}}},  \label{time}
\end{equation}%
With these new variables, eqs. (\ref{kinetic}) take the form: 
\begin{eqnarray}
\frac{dx_i}{dz}=\frac{1}{H_0 \left( 1+z \right)\sqrt{\Omega_{M,0}(1+z)^3 +
\Omega_{\Lambda}}}\times \cr \left[ n_{0}\left( 1+z\right) ^{3}\left(
\sum_{j,k}x_{j}x_{k}R_{jki}-x_{i}\sum_{m,n}x_{m}R_{imn}\right)\right. \cr %
\left. + \sum_{j}x_{j}R_{ji} - x_{i}\sum_{m}R_{im}\right].  \label{17}
\end{eqnarray}

As was mentioned above, the reaction rates $R_{ijk}$ and $R_{ij}$ are
functions of the radiative temperature $T_{r}$ and the kinetic temperature $%
T_{k}$, which depend on many parameters. The radiative temperature can be
taken as in the case of the adiabatic expansion, $T_{r}=T_{0}\left(
1+z\right) $, but in order to find the kinetic temperature we have to solve the
differential equation for $T_{k}$ : 
\begin{eqnarray}
\frac{dt_{k}}{dt}=-2T_{k}H_{0}\sqrt{\Omega _{M,0}(1+z)^{3}+\Omega _{\Lambda }%
}+\left( \frac{dt_{k}}{dt}\right) _{mol}+ \cr \frac{8\sigma _{t}a_{b}T_{r}^{4}%
}{3m_{e}c}x_{e}\left( T_{r}-T_{k}\right), 
\end{eqnarray}
where the first term in the right--hand part describes the change on the
temperature due to the universe expansion, the second one corresponds to the
thermodynamics of the primordial molecules, and the third one is Thomson
scattering. The second term can be written as follows: 
\begin{equation}
\left( \frac{dt_{k}}{dt}\right) _{mol}=\frac{2\left( \Gamma _{mol}-\Lambda
_{mol}\right) }{3nk}+\frac{\Theta _{ch}}{3nk}+\frac{2T_{k}}{3n}\left( \frac{%
dn}{dt}\right) _{ch},  \label{secondterm}
\end{equation}%
where $\Gamma _{mol}$ and $\Lambda _{mol}$ are the heating and cooling
functions, $\Theta _{ch}$ is the energy gain (loss) due to the chemical
reactions, characterized by their enthalpy $\Delta H$, 
\begin{equation}
\Theta _{ch}=\sum_{reactions}R_{ijk}n_{i}n_{j}\Delta H_{,}  \label{enthalpy}
\end{equation}%
The summation is carried out over all chemical reactions.

The factor $\left( \frac{dn}{dt}\right) _{ch}$ in the last term of eq. (\ref%
{secondterm}) corresponds to the change in the primordial gas density due to
the chemical reactions, which do not conserve the number of initial species
(for example $A+B\rightarrow C$, or $A+B\rightarrow C+D+E$).

The calculations of the molecular abundances were carried out for a wide
range of redshifts for the following species involved in HD molecule
formation: H, H$^{+}$, H$_{2}^{+}$, H$_{2}$, H$^{-}$, e$^{-}$, H$_{3}^{+}$,
He, He$^{+}$, HeH$^{+}$, D, HD, H$_{2}$D$^{+}$, D$^{+}$, D$^{-}$, and HD$%
^{+} $. Other species are negligible in the formation processes of the HD
molecule due to their small abundances. 
The following reactions are of particular importance in the
formation/destruction processes of the HD molecule in the primordial 
gas:
\begin{eqnarray} 
&&{\rm D + H_2 = HD + H} \nonumber \\
&&{\rm D^+ + H_2 = HD + H^+} \nonumber  \\
&&{\rm D^- + H = HD + e^-} \nonumber \\
&&{\rm H^- + D = HD + e^-} \nonumber \\
&&{\rm H + HD = H_2 + D} \nonumber \\
&&{\rm H^+ + HD = H_2 + D^+} 
\end{eqnarray}
More detailed consideration can be found in the many papers (see for 
example Galli \& Palla 1998,2002). 
The results of our molecular abundance calculations are shown in Fig. 1. One
can see that the relative abundance ratio [HD]/[H$_{2}$] at low redshifts is
slightly larger than $10^{-3}$, whereas at high redshifts this value tends
to the assumed primordial ratio [D]/[H]=3 \ 10$^{-5}$. This fact is explained 
as the deuteration process, D$^{+}+$H$_{2}\ \rightarrow $ H$^{+}+$HD 
and D $+$ H$_2 \rightarrow$ HD $+$ H, that leads to
relatively high abundance of the molecule, up to [HD]/[H$_{2}$]$=10^{-2.39}$
at $z=10$ in collapsed clouds (see e.g., Galli \& Palla 2002).

\begin{figure}[tbp]
\vspace{7cm} \includegraphics{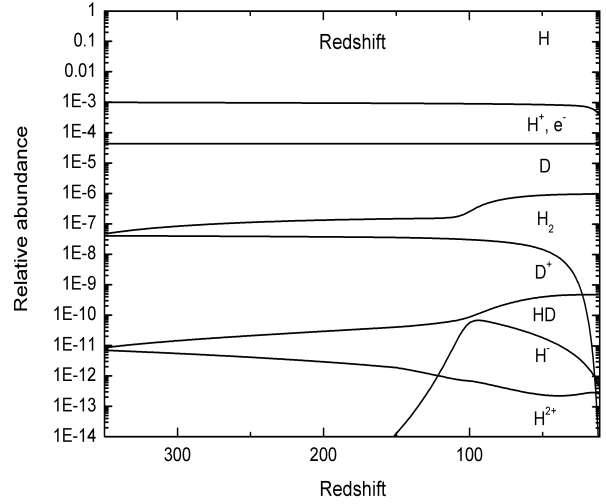}
\caption{Fractional abundance of chemical species involved in the HD
molecule formation as a function of redshift for the expanding
and adiabatically cooling homogeneous primordial medium. The concordance 
$\Lambda$CDM cosmological model was used.}
\end{figure}


\section{The HD optical depth in protohaloes and haloes at high redshifts}


We are now interested in obtaining estimates related to the observability of
the line features of the primordial HD molecule in protostructures in their
linear or quasi--linear evolution regime, i.e. before gravitational collapse,
and first star formation.
As mentioned in the Introduction, previous works have shown that the largest
resonant line opacities in homogeneous mass perturbations are produced when
they are at their maximum expansion or turn--around epoch. Based on these
conclusions, here we will calculate the HD resonant line opacities at
the turnaround epoch of CDM high--peak mass perturbations. After the
first bursts of star formation (Population III stars) in the high--peak ($%
3-6\sigma $) collapsed perturbations, re--ionization, feedback and metal
enrichment processes change drastically the properties of the medium in the
lower density mass perturbations. This is why we will focus on
the protostructures emerging from high 3--6$\sigma$ peaks. The interaction
of CMBR photons with the primordial molecules inside these protostructures
will produce signatures related to the pristine conditions of gas before
the first star formation in the universe.

\subsection{Properties of 3 and $6\sigma$ top--hat CDM protohaloes and haloes}
\begin{table*}\centering
  \caption{Properties of $\Lambda$CDM mass perturbations}
  \begin{tabular}{@{}lrlllrlllrrl@{}}
  \hline
   $M$ & $z_{ta}$\hspace{2mm} & $\rho_{ta}$ & $n_{ta}$ & $R_{ta}$ & $z_{col}\hspace{1.5mm}$
     & $\rho_{col}$ & $n_{col}$ & $R_{col}$ & $t_{dyn} \hspace{2.5mm}$ & $z_{cool}\hspace{0.5mm}$ 
     & $n_{cool}$ \\
   $(M_\odot)$ & & $(gcm^{-3})$ & $(cm^{-3})$ & $(cm)$ & & $(gcm^{-3})$ & $(cm^{-3})$ & $(cm)$ & $(Myr)$ & & $ (cm^{-3})$ \\
 \hline
 \multicolumn{12}{c}{Case $3\sigma$ } \\
$10^{2}$ & 58.52 & 3.02 $10^{-24}$ & 2.67 $10^{-1}$ & 2.50 $10^{19}$ & 36.49 & 2.40 $10^{-23}$ & 2.12 $10^{+0}$  & 1.26 $10^{19}$ & 13.58 & 32.65 & 2.12 $10^{3}$ \\
$10^{4}$ & 46.26 & 1.51 $10^{-24}$ & 1.34 $10^{-1}$ & 1.46 $10^{20}$ & 28.77 & 1.20 $10^{-23}$ & 1.06 $10^{+0}$  & 7.34 $10^{19}$ & 19.20 & 25.72 & 1.06 $10^{3}$ \\
$10^{6}$ & 34.55 & 6.44 $10^{-25}$ & 5.70 $10^{-2}$ & 9.03 $10^{20}$ & 21.40 & 5.11 $10^{-24}$ & 4.52 $10^{-1}$  & 4.53 $10^{20}$ & 29.43 & 19.10 & 4.52 $10^{2}$ \\
$10^{8}$ & 24.17 & 2.28 $10^{-25}$ & 2.02 $10^{-2}$ & 5.92 $10^{21}$ & 14.85 & 1.81 $10^{-24}$ & 1.60 $10^{-1}$  & 2.97 $10^{21}$ & 49.40 & 13.23 & 1.60 $10^{2}$ \\
$10^{10}$ & 15.18 & 6.07 $10^{-26}$ & 5.37 $10^{-3}$ & 4.28 $10^{22}$ & 9.19 & 4.82 $10^{-25}$ & 4.27 $10^{-2}$  & 2.14 $10^{22}$ & 95.78 & 8.15 & 4.27 $10^{1}$ \\
$10^{12}$ & 7.92 & 1.02 $10^{-26}$ & 8.99 $10^{-4}$ & 3.60 $10^{23}$ & 4.61 & 8.10 $10^{-26}$ & 7.16 $10^{-3}$  & 1.80 $10^{23}$ & 233.69 & 4.03 & 7.16 $10^{0}$ \\
\hline
\multicolumn{12}{c}{Case $6\sigma$ } \\
$10^{2}$ & 118.04 & 2.42 $10^{-23}$ & 2.14 $10^{+0}$ & 1.25 $10^{19}$ & 73.99 & 1.92 $10^{-22}$ & 1.70 $10^{+1}$  & 6.28 $10^{18}$ &4.80 & 66.30 & 1.70 $10^{4}$ \\
$10^{4}$ & 93.53 & 1.21 $10^{-23}$ & 1.07 $10^{+0}$ & 7.32 $10^{19}$ & 58.55 & 9.60 $10^{-23}$ & 8.49 $10^{+0}$  & 3.67 $10^{19}$ & 6.79 & 52.45 & 8.49 $10^{3}$ \\
$10^{6}$ & 70.10 & 5.15 $10^{-24}$ & 4.56 $10^{-1}$ & 4.52 $10^{20}$ & 43.79 & 4.08 $10^{-23}$ & 3.61 $10^{+0}$  & 2.26 $10^{20}$ & 10.40 & 39.20 & 3.61 $10^{3}$ \\
$10^{8}$ & 49.34 & 1.83 $10^{-24}$ & 1.62 $10^{-1}$ & 2.96 $10^{21}$ & 30.71 & 1.45 $10^{-23}$ & 1.28 $10^{+0}$  & 1.48 $10^{21}$ & 17.47 & 27.46 & 1.28 $10^{3}$ \\
$10^{10}$ & 31.37 & 4.86 $10^{-25}$ & 4.30 $10^{-2}$ & 2.14 $10^{22}$ & 19.39 & 3.86 $10^{-24}$ & 3.41 $10^{-1}$  & 1.07 $10^{22}$ & 33.86 & 17.30 & 3.41 $10^{2}$ \\
$10^{12}$ & 16.86 & 8.16 $10^{-26}$ & 7.22 $10^{-3}$ & 1.80 $10^{23}$ & 10.25 & 6.48 $10^{-25}$ & 5.73 $10^{-2}$  & 9.02 $10^{22}$ & 82.62 & 9.10 & 5.73 $10^{1}$ \\
\hline
\end{tabular}%
\end{table*}

To calculate the optical depth of primordial protoclouds by scattering of
CMBR photons on HD molecules we need a model for these protoclouds. We
use the spherical top--hat approach to calculate the epoch, density, and size
of CDM mass perturbations at their maximum expansion for the $\Lambda$CDM
cosmology adopted here. This approach is justified for the kind of estimates
in which we are interested (e.g., Tegmark et al. 1997), and for the
quasi--lineal phases of gravitational evolution. It is assumed that gas is
equally distributed as CDM in the mass perturbation but with a density $%
\Omega_{B,0}/\Omega_{M,0}$ times lower.
According to the spherical top--hat approach, the overdensity of a given mass
perturbation, $\delta (M)$, grows with $z$ proportional to the so--called
growing factor, $D(z)$, until it reaches a (linearly extrapolated) critical
value, $\delta _{c}$, after which the perturbation is supposed to collapse
and virialize at redshift $z_{\rm col}$ (for example see Padmanabhan 1993): 
\begin{equation}
\delta (M;z_{\mathrm{col}})\equiv \delta _{0}(M)D(z_{\mathrm{col}})=\delta
_{c,0}.  \label{delta}
\end{equation}%
The convention is to fix all the quantities to their linearly extrapolated
values at the present epoch (indicated by the subscript ``0'') in such a way
that $D(z=0)\equiv D_0=1$. 

The redshift of turn--around, 
$z_{\mathrm{ta}}$, is calcualted by using the top--hat sphere result of 
$t_{\mathrm{ta}}=0.5t_{\mathrm{col}}$.
Following the evolution equation for the top--hat sphere, one finds for our
flat $\Lambda$CDM cosmology that the average sphere density at $z_{\mathrm{ta}}$ 
is given by $\rho _{\mathrm{ta}}\approx 5.6\rho _{\mathrm{bg}}(z_{\mathrm{ta}})$%
, i.e. similar to the Einstein--de Sitter case (Wang \& Steinhardt 1998;
Horellou \& Berge 2005). The background density at $z_{\mathrm{ta}}$ is $%
\rho _{\mathrm{bg}}(z_{\mathrm{ta}})=1.88\ 10^{-29}$g cm$^{-3}\Omega
_{M,0}h^{2}(1+z_{\mathrm{ta}})^{3}$. The corresponding numerical gas density
is $n_{\mathrm{ta}}=(\rho _{\mathrm{ta}}/m_{p})(\Omega_{B,0}/\Omega_{M,0})$, 
where $m_{p}$ is the H atom mass. Finally, the radius of the sphere of mass $M$ 
at the turn--around or maximum expansion is given by 
\begin{equation}
R_{\mathrm{ta}}=\left[ \frac{3M}{4\pi \rho _{\mathrm{ta}}}\right] ^{\frac{1}{%
3}}.  \label{radmax}
\end{equation}

We need now to connect the top--hat sphere results to a characteristic mass
of the perturbation, $M$, produced within the $\Lambda$CDM structure formation 
scenario. The primordial fluctuation field, fixed at the present epoch, is 
characterized by the mass variance, $\sigma(M)$, which is the rms mass perturbation
smoothed on a given scale $R$ corresponding to the mass $M$.
For our calculations, we use the
power spectrum shape given by Bardeen et al. (1986) normalized to $%
\sigma_8=0.9$. Thus, the mass perturbation
overdensity linearly extrapolated to $z=0$ can be defined as: 
\begin{equation}
\delta_0(M) = \nu\sigma(M),  \label{d0CDM}
\end{equation}
where  $M$ and $\nu$ are the perturbation mass and peak height, respectively. 
For average perturbations, $\nu=1$, while for rare, high--density perturbations, 
from which emerged the first structures, $\nu>>1$.

By introducing eq. (\ref{d0CDM}) into eq. (\ref{delta}) one may infer $z_{%
\mathrm{col}}$, and therefore $z_{\mathrm{ta}}$, $\rho _{\mathrm{ta }}$, $n_{%
\mathrm{ta }}$, and $R_{\mathrm{ta}}$ for $\nu\sigma$ mass perturbations
from the $\Lambda$CDM primordial fluctuation field. In Table 1 we present
these quantities for $3\sigma$ perturbations of different masses. As
mentioned above, we are interested in HD line signatures from the
protoclouds before first luminous objects in the universe formed and started
to re--ionize it, and therefore only high--peak low mass perturbations will 
be considered.

Cosmological N--body + hydrodynamics simulations showed that the first stars
may have formed indeed at redshifts as high as $\sim 18-30$ in ($\sim 3-4\sigma$)
CDM mini--haloes of M$\sim 10^6\msun$ (Abell et al. 1998; 2002; Fuller \& Couchman
2000; Bromm et al. 2002; Yoshida et al. 2003). Related to this is the fact
that haloes of any given mass that collapse first (high peaks) do not
populate ``typical'' regions at all, but rather ``protocluster'' regions
(White \& Springel 2000; Barkana \& Loeb 2002). In a recent paper, Gao et
al. (2005) used a novel technique to follow with unprecedentedly high resolution
the growth of the most massive progenitor of a supercluster region from $%
z\approx 80$ to $z=0$. By using this simulation, Reed et al. (2005) have
found that, when the mass of the progenitor halo is $\approx 2.4 \ 10^{5}$h$%
^{-1}\msun$ at $z\approx 47$, it undergoes baryonic collapse via H$_2$
cooling, triggering star formation at this early epoch. Of course, this halo
should emerge from a very rare fluctuation peak, $\nu=6.5$ (Reed et al. 2005). 
Notice that rare peaks are strongly clustered, so that the probability
to find a cluster of these (proto)halos is high.
In the second part of Table 1 we present the same as in the first part, 
but for $6\sigma$ haloes.

\subsection{Optical depths}

The optical depth for a protocloud of size $L$ is given by: 
\begin{equation}
\tau _{\nu }\left( L\right) =\int_{0}^{L}\alpha _{\nu }\left( x\right) dx,
\label{21}
\end{equation}%
where $\alpha _{\nu }\left( x\right) $\ is the absorption coefficient, and
integration is carried out over the line of sight. In the case of a mass
perturbation at its turn--around epoch, when the physical parameters of the
gas are approximately equal for all parts of the spherical region, the
integration can be reduced to a more simple expression: $\tau _{\nu }\left(
L\right) \approx \alpha _{\nu }L$. So, the optical depth in this case can be
written as: 
\begin{equation}
\tau _{\nu }=\frac{\lambda ^{3}\left( 2J^{\prime }+1\right) }{8\pi \left(
2J+1\right) V_{T}}x_{\mathrm{HD}}n_{\mathrm{J}}n_{B}A_{J^{^{\prime
}}J}\left( 1-e^{\frac{-h\nu }{\kappa T_{r}}}\right) L,  \label{tau2}
\end{equation}%
where $\lambda $\ is the wavelength, $J$\ is the rotational quantum number, $%
x_{\mathrm{HD}}$\ is the relative abundance of the HD molecule, 
$A_{J^{^{\prime }}J}$\ is the Einstein coefficient, and $n_B$ and
$n_{\mathrm{J}}$\ are the baryon and $J$--th rotational level population 
numerical densities at epoch $z$, respectively. As to $n_{J}\left(
z\right) $, it is well known that at the low density limit ($n_{H}<10^{3}$cm$%
^{-3}$), the population differs from the Boltzmann one. So, in the general
case, when the gas is out of thermodynamical equilibrium, the correct values
of the ro--vibrational levels population should be calculated with the
detailed balance equation: 
\begin{eqnarray}
n_{vJ}\sum_{v^{\prime }J^{\prime }}(W_{vJ\rightarrow v^{\prime }J^{\prime
}}^{R}+W_{vJ\rightarrow v^{\prime }J^{\prime }}^{C})= \cr \sum_{v^{\prime
}J^{\prime }}n_{v^{\prime }J^{\prime }}(W_{v^{\prime }J^{\prime }\rightarrow
vJ}^{R}+W_{v^{\prime }J^{\prime }\rightarrow vJ}^{C}),  \label{pop}
\end{eqnarray}%
where $n_{vJ}$\ is the population of the ro--vibrational level $vJ$, $%
W_{vJ\rightarrow v^{\prime }J^{\prime }}^{R}$\ and $W_{vJ\rightarrow
v^{\prime }J^{\prime }}^{C}$\ are the probabilities of the radiative and
collisional transitions, respectively.

By using the expressions (\ref{tau2}) and (\ref{pop}), we calculate the HD
line optical depths corresponding to protohaloes of several masses at their
turnaround redshifts (Table 1). The results corresponding to three ground 
rotational line transitions (1--0), (2--1) and (3--2) for the $3\sigma$ and 
$6\sigma$ protohaloes are plotted in Figs. 2 and 3, respectively. As one can
see, the values of the optical depth for the HD molecule lines are rather large
as compared with those reported for the LiH molecule (Bougleux \& Galli 1997).
Even the transition $J^{\prime}-J=3-2$ reaches values as high as 
$10^{-9}-10^{-8}$ for the redshifts $z\approx 20-40$.

The estimates presented above correspond to the linear evolution phase of
fluctuations. The calculation of chemical kinematics evolution of HD
molecule during the gas cooling and collapse inside the virializing dark
haloes is very difficult since it is linked to the thermal and 
hydrodynamical evolution. Under several assumptions and simplifications,
Galli \& Palla (2002) presented results of HD abundances after the gas cooling
inside virialized CDM haloes constructed according to Tegmark et al. (1997).
The result of Galli \& Palla is that the HD relative fraction, $x_{HD}$, increases
dramatically, approximately by a factor of 200, with respect to the initial
one at the collapse epoch of the halo. In the following, we will use this result to
get a rough estimate of the HD relative fraction in collapsed $\Lambda$CDM
haloes and the corresponding $\tau_\nu$.

We have calculated above the evolution of the HD relative fraction, $x_{HD}$, 
in the homogeneous pregalactic medium. For a given mass fluctuation $%
\nu\sigma $, we have obtained $x_{\mathrm{HD}}$ at $z_{\mathrm{ta}}$. Let us
assume that after the halo collapsed, at $z_{\mathrm{col}}$, $n_{HD}$
increased only proportionally to the halo density increasing. Therefore, $n_{%
\mathrm{HD}}(z_{\mathrm{col}})=n_{\mathrm{HD}}(z_{\mathrm{ta}})\times \rho _{%
\mathrm{col}}/\rho _{\mathrm{ta}}$. Then, we assume that the gas cools
rapidly and falls to the halo center in a dynamical time, 
\begin{equation}
t_{\mathrm{dyn}}\approx \frac{\pi /2}{\sqrt{2GM/R_{\mathrm{col}}^{3}}}
\end{equation}%
where $R_{\mathrm{col}}$ is the halo radius. 
Now, according to the Galli \& Palla result, we assume that at the epoch when
the gas cooled and collapsed, $z_{\mathrm{cool}}$, the HD relative fraction
increased by a factor of $\sim 200$, i.e. 
$x_{\rm HD}(z_{\rm cool})\approx 200\ x_{\rm {HD}}(z_{\rm {col}})$. The collapsing
factor of the gas is assumed to be 10, so that the number density increases
by a factor of 1000.
We further assume that the kinetic temperature of the cooled gas is equal to the 
CMBR radiation (see Fig. 10 in Galli \& Palla).  The populations corresponding
to the main ground line transition ($J^\prime-J$)=(1--0) actually depend weakly 
on $T_k$. In Fig. 4 we show the HD ($J^\prime-J$)=(1--0)
line transition optical depths corresponding to 3 and $6\sigma$ haloes of several
masses at their ``cooling'' redshifts. The radii used to calculate $\tau_{\nu }$ 
were 1/10th of the corresponding halo radii $R_{\rm {col}}$.
For the redshifts of interest, i.e. before massive star formation ($z\gsim 20$),
the optical depths of the lines are significantly high, with values
from several times $10^{-2}$ to several times  $10^{-5}$, up to $z\sim 50$. 

\begin{figure}[tbp]
\vspace{6cm} \includegraphics{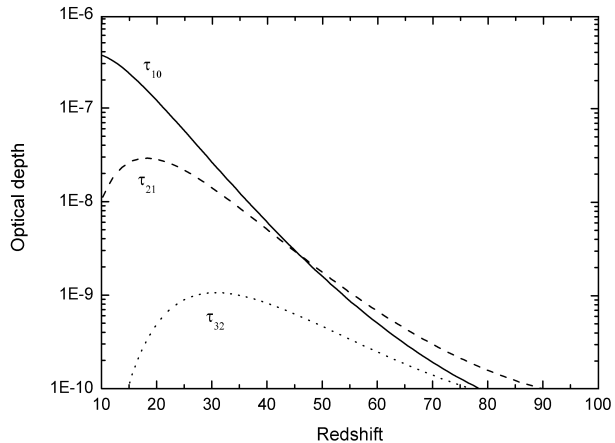}
\caption{Optical depths in the first, second and third rotational
transitions of HD molecule in $3\sigma$ $\Lambda $CDM protohaloes of 
different masses reaching their maximum expansion at the redshifts shown 
in the abscissa.}
\end{figure}

\begin{figure}[tbp]
\vspace{6cm} \includegraphics{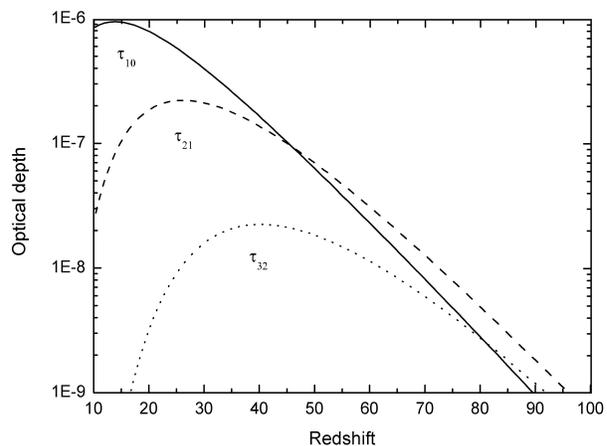}
\caption{Same as in Fig. 2 but for $6\sigma $ $\Lambda $CDM 
protohaloes.}
\end{figure}

\begin{figure}[tbp]
\vspace{6cm} \includegraphics{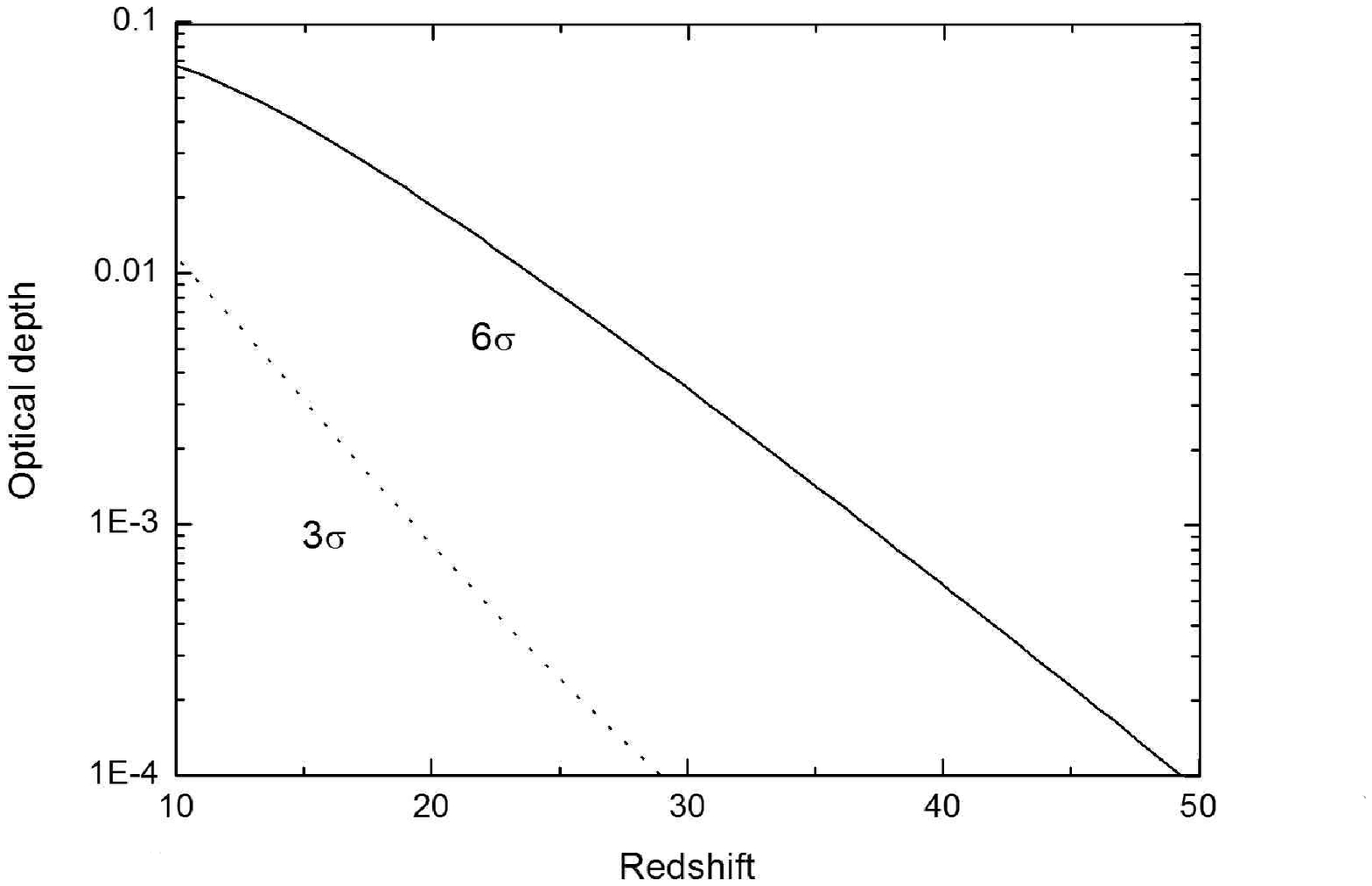}
\caption{Same as in Figs. 2 and 3 but for haloes already collapsed and at 
the corresponding redshifts when the baryons have supposedly cooled and collapsed 
inside them (see text). Only the case for the first rotational transition
is showed.}
\end{figure}

\subsection{Observational estimates}

As shown by Sunyaev \& Zel'dovich (1970) (see also Dubrovich 1977;
Zel'dovich 1978; Maoli et al. 1994) the amplitude of 
secondary anisotropies in the CMBR temperature due to Doppler effect 
of resonance scattered photons in protoclouds with peculiar motions is: 
\begin{equation}
\frac{\Delta T}{T}=\frac{V_{p}}{c}\tau_\nu  \label{deltaT}
\end{equation}%
for $\tau_\nu<<1$, where $V_{p}=V_{p}(z)$ is the peculiar velocity of the 
protocloud with respect to the CMBR at epoch $z$, $c$ is the light speed, 
and $\tau_\nu $\ is the optical depth at frequency $\nu$ through the 
protocloud at epoch $z$ (to be the protocloud $z_{\mathrm{ta}}$ in our 
analysis). The linear theory of gravitational instability shows that the
peculiar velocity of every mass element grows with the expansion factor as $%
V_{p}(z)\propto \dot{D}(z)/(1+z)$. An accurate approximation to this
expression for a flat universe with cosmological constant is (Lahav et al.
1991; Carroll et al. 1992): 
\begin{equation}
V_{p}(z)\propto f(z)\frac{g(z)}{g(0)}\frac{E(z)}{(1+z)^{2}},  
\label{vpec}
\end{equation}%
where $f(z)\approx \Omega _{M}(z)^{0.6}$, and the functions $g(z)$, $E(z)$
and $\Omega _{M}(z)$ were defined in \S 3.1. For the Einstein--de Sitter
universe, eq. (\ref{vpec}) reduces to the known expression $V_{p}(z)\propto
(1+z)^{-2}$. 

The peculiar velocity field at $z=0$, traced by galaxies and galaxy
clusters, has been measured in a large range of scales (for a review see
e.g., Zaroubi 2002). For our problem, the velocity field traced by galaxy
clusters is probably the interesting one because, as mentioned above, the
first luminous objects in the universe formed in the densest, most rare
regions, i.e. those that today are clusters of galaxies. Various data sets
lead to different rms peculiar velocities for scales $\gsim 100$h$^{-1}$Mpc.
Here we will adopt a value of $V_{p}(0)\approx 650$km/s, in agreement with
results from Lauer \& Potsman (1994), Hudson et al. (1999,2004), and Willick
(1999). This value is apparently a factor of $\sim 2$ larger than the one
corresponding to $\Lambda$CDM numerical simulations for the scales of
interest.

By using eq. (\ref{vpec}) normalized to $V_{p}(0)=650$km/s and the HD
optical depths calculated in \S 3.2, we may estimate with eq. (\ref{deltaT})
the corresponding temperature fluctuation of the secondary anisotropies. In
the range of $z=20-40$, for the case of the $3\sigma $ protohaloes, we obtain
that $\Delta T/T\approx \ 10^{-10}-4\ 10^{-12}$ in the first rotational
transition, while for the $6\sigma $ protohaloes, $\Delta T/T\approx 5$ $%
10^{-10}-10^{-10}$. In the same redshift range, for the collapsed baryonic 
clouds, whose HD abundances were estimated using results from Galli \& Palla 
(2002), we calculate $\Delta T/T\approx \ 10^{-6}-10^{-8}$ and
$\Delta T/T\approx \ 2\ 10^{-5}-10^{-6}$ for the 3 and $6\sigma$ cases,
respectively. 

Now let us estimate the integration time required for the detection of
the secondary anisotropies with millimeter and submillimeter
facilities under construction as for example: the 50m LMT/GTM telescope in 
Sierra Negra, Mexico; the ``Combined Array for Research in Millimeter--Wave 
Astronomy'' (CARMA) in USA; the ``Atacama Large Millimeter Array'' (ALMA) in 
Chile. The observational (integration) time $\Delta t$\ can be estimated from 
the equation 
\begin{equation}
\Delta T=\frac{T_{n}}{\sqrt{\Delta \nu \Delta t}},  
\label{dt}
\end{equation}%
where $\Delta \nu $\ is the bandwidth, $T_{n}$ is the detector noise temperature 
and $\Delta T$\ is the amplitude of the temperature fluctuation calculated above.
For the three facilities mentioned, $T_n\approx 40-80$ K, and we may assume 
$\Delta \nu \approx 1$. In the case of GTM/LMT telescope, its angular 
resolution limits the observability of several cases (see below). By 
using eq. (\ref{dt}), for the $3\sigma $ and $6\sigma $ protohaloes, we 
find that the integration times
required to get an observable signal of the first rotational transition of
HD molecules are too large to be observed. In the case of collapsed gas clouds 
(previous to star formation triggering) inside virialized $6\sigma$ haloes, 
we estimate $\Delta t \lsim 10^{5}s = 28\ $\ hours for $z < 40$ and
$T_n\approx 50$ K. This integration time can be attained at different observational 
sessions of several hours each. 

Let us discuss another important parameter of the potentially observed 
protoclouds in molecular resonant line emission: their apparent angular 
size or diameter. Apparent angular size for an object of the 
linear size $L$ is given by the Hoyle's formula: 
\begin{equation}
\Delta \theta =\frac{L}{d_{A}(z)},  \label{Hoyle}
\end{equation}%
where $d_{A}(z)$ is the angular-diameter distance.

The angular--diameter distance for a flat universe with cosmological
constant is given by: 
\begin{equation}
d_{A}(z)=\frac{c}{\left( 1+z\right)} \int_{0}^{z}\frac{dz^{\prime }}{H(z^{\prime })},
\label{dL}
\end{equation}%
where the Hubble parameter in this case is: 
\begin{equation}
H(z)=H_{0}\sqrt{\Omega _{M,0}(1+z)^{3}+\Omega _{\Lambda }},  \label{29}
\end{equation}%
and the radiation density term was neglected. Now eq. (\ref{Hoyle}) can
be written as follows: 
\begin{equation}
\Delta \theta =\frac{H_{0}L(1+z)}{c\int_{0}^{Z}\left( \Omega
_{M,0}(1+z^{\prime })^{3}+\Omega _{\Lambda }\right) ^{-1/2}dz^{\prime }},
\label{angsize}
\end{equation}

We use eq. (\ref{angsize}) to calculate the angular size $\Delta \theta $ of 
$3$ and $6\sigma $ protoclouds reaching their maximum expansion in the
redshift range of $20<z<40$ (corresponding roughly to mass ranges of $%
10^{5}<M/\msun<10^{9}$ and $10^{9}<M/\msun<10^{11}$ for the $3\sigma $ and $%
6\sigma$ cases, respectively). We calculate also the corresponding
redshifted frequencies, $\nu _{0}$, for the same three lowest rotational
transitions of HD molecule considered in Figs. 2 and 3. Our results are shown in
Fig. 5, along with the expected coverage in this $\nu _{0}-\Delta \theta $
plane for the LMT/GTM and CARMA facilities (boxes). The ALMA facility covers
the whole plotted plane. We also include in this plot 
the curves corresponding to the cold clouds within collapsed $3$ and $6\sigma$ 
haloes for only the first rotational transition line (thin lines with square dots).
These clouds could be resolved by ALMA and only partially by CARMA.
It should be noted that high--peak halos are strongly clustered
(Gao et al. 2005; Reed et al. 2005). It may thus be more plausible
to find a cluster of such objects in a patch of the sky even if individual 
objects are very small. Moreover, the clustering increases the amplitude
of the spectra--spatial fluctuations in the CMBR temperature.
Figure 5 is rather general and can be applied to any other resonant lines 
of similar frequencies in which the gas in protohaloes or haloes could be detected.

\begin{figure}[tbp]
\vspace{6.5cm} \includegraphics{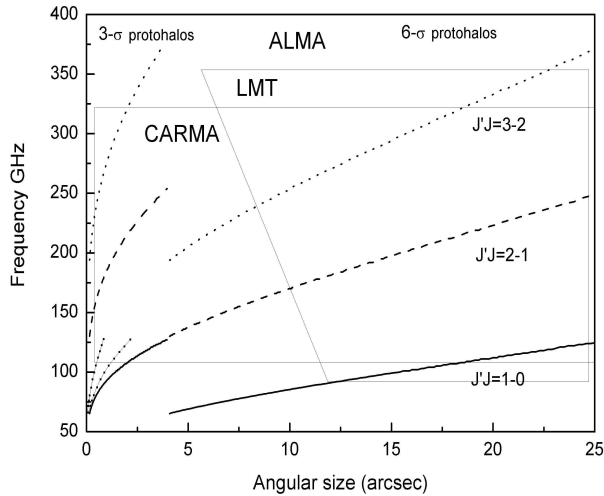}
\caption{Relation between the angular size $\Delta \protect\theta $ of the $3%
\protect\sigma $ and $6\protect\sigma $ protoclouds reaching their maximum
expansion at $20<z<40$ and the redshifted frequency for three lowest
rotational transitions of the HD molecule. Thin lines with squared dots are 
for $3\sigma$ (upper curve) and $6\sigma$ (lower curve) collapsed cold 
clouds in the same redshift range and only for the lowest transition. The 
boxes indicate qualitatively the regions in which the LMT/GTM and CARMA will 
operate. ALMA will cover almost the whole plotted plane.}
\end{figure}

\section{Discussion and Conclusions}

The hope to detect signals from the end of the so--called 'dark age'
has prompted a flurry of theoretical and instrumental activity. The
observation of high--redshift protostructures before they 
collapse and star formation triggers will certainly help us to 
constrain structure formation scenarios. One possibility of such
observations is based on the spectral--spatial fluctuations in the CMBR 
temperature produced by elastic resonant scattering 
of CMBR photons on molecules located in protostructures moving with peculiar 
velocity. This kind of secondary anisotropies search for LiH 
molecule has been fruitless (de Bernardis et al. 1993). As Bougleux
\& Galli (1997) showed, the abundances of LiH molecule and their optical 
depths in the rotational lines are too small to produce detectable
CMBR temperature fluctuations. 

In this paper we have investigated the spectral--spatial fluctuations 
in the the CMBR temperature due to elastic resonant scattering by HD molecules. 
For this molecule, the main contribution to the optical depth comes from the 
ground rotational transition ($J^\prime-J$)=(1--0), because of small 
populations with
high rotational levels of the HD molecule at early epochs. In the case
of the LiH molecule, most of the contribution to the optical depth comes from
rotational transitions between high $J$ levels.  The wavelenght at rest of 
the HD ($J^\prime-J$)=(1--0) transition is 112.1 $\mu m$. Therefore, for $z=20-40$ 
it should be observed at the wavelengths $\approx 2-4.5$mm. For higher
rotational transitions, the wavelengths are smaller. 

We have carried out calculations in order to estimate the HD optical 
depths and the corresponding amplitudes of the spectral--spatial 
fluctuations in the CMBR temperature produced by moving protostructures. First 
we calculated detailed chemical kinematic evolution for HD molecule in the 
expanding and adiabatically cooling homogeneous medium, assuming standard
BBN yields.  We further calculated
the optical depths in HD pure rotational lines for mass perturbations 
(protostructures) at their
maximum expansion under the assumption that the HD relative fractions, 
$x_{HD}$, within these protostructures are similar to the fractions
calculated for the homogeneous medium at a given $z$. The spherical top--hat
approach and the concordance $\Lambda$CDM cosmology were used to calculate
the corresponding turn--around redshifts, sizes, and peculiar velocities
for different masses and peaks. Once the optical depths for the first three 
ground rotational transitions were calculated, the corresponding fluctuations 
$\Delta T/T$ were estimated. We also estimated $\tau_{\nu}$ and $\Delta T/T$ 
for cold gas clouds inside collapsed CDM haloes, but under several 
assumptions and using results on the HD relative fraction from a crude
calculation by Galli \& Palla (2002).
    
The main results from our study are the following: 

$\bullet$ The relative fraction of HD molecule, $x_{\rm HD}$, increases in the 
homogeneous expanding medium by a factor of $\approx 400$ since $z=300$ to low 
redshifts (Fig. 1). At redshifts $40-20$, $x_{\rm HD}\approx 4-5\ 10^{-10}$.
These fractions are expected to strongly increase during the cooling and
collapse of the gas within CDM haloes. 

$\bullet$ The optical depth of the HD first pure rotational line in high--peak 
protostructures at their maximum expansion increases with time.
For the second and third lines, the corresponding $\tau_\nu$'s attain
a maximum at high redshifts and then strongly decrease for lower redshifts. 
The ranges of $\tau_\nu$ values of the (1--0) transition lines in 3 and $6\sigma$ 
protostructures at their maximum expansion in the $20<z<40$ redshift interval are 
approximately 
$10^{-7}-6\ 10^{-9}$ and $8\ 10^{-7}-2\ 10^{-7}$, respectively (Figs. 2 and 3).
For the crude model of cold gas clouds inside collapsed 3 and $6\sigma$ CDM haloes,
the ranges of $\tau_\nu$ in the same redshift interval are $8\ 10^{-4}-10^{-5}$
and $2\ 10^{-2}-1.3\ 10^{-3}$, 
respectively (Fig. 4). The optical depths of HD rotational lines in 
protostructures are much higher than those of the LiH molecule. 

$\bullet$ The ranges of redshifted frequencies and typical angular sizes in the sky
of spectral--spatial fluctuations due to HD molecule rotational lines
in high--peak protostructures and gas clouds in collapsed haloes fall partially 
within the observational windows of submillimeter telescope facilities under 
construction, as 
GTM/LMT, CARMA and ALMA (Fig. 5). The range of redshifts under consideration was
$20<z<40$; after this epoch the first baryonic structures are probably
already in site with strong emitting stellar sources. The observational search 
of the spectral--spatial fluctuations studied here will be important for testing 
models of cosmic structure formation as well as the primordial abundance of
deuterium predicted in standard and non-standard BBN theories. 

$\bullet$ For the $\Lambda$CDM scenario, the amplitudes of the spectral--spatial 
fluctuations produced by HD molecule at $20<z<40$ in protostructures, even 
as rare as $3-6\sigma$ density--peaks, are too faint for reaching the flux 
sensitivity 
of submillimeter/millimeter facilities under construction. For the 
case of cold gas clouds within collapsed $\Lambda$CDM haloes emerging from 
$\sim 6\sigma$ density--peaks, the estimated amplitudes, in particular for the 
lowest rotational transition line, are much larger and could be detected, but 
their estimated angular sizes in the sky are too small for telescopes such as 
GTM/LMT, but possibly clusters of such objects could be resolved (high--peak
halos are strongly clustered).
The spectral--spatial fluctuations predicted
for very rare high--redshift cold clouds in the $\Lambda$CDM scenario will 
fall within the observational capabilities of other facilities as CARMA and ALMA.
For non-standard BBN theories, the [D/H] abundance can be much larger than
the used here. For example, in inhomegeneous BBN models, [D/H] is estimated
to be ten times larger than in the standard BBN (e.g., Lara 2005). Therefore
the HD line optical depths presented here could be a factor of ten higher,
making already detectable the corresponding spectral--spatial fluctuations
produced at the maximum expansion of protohaloes. 

A more detailed study of spectral--spatial fluctuations due to HD molecules in 
collapsing gas clouds as well as the inclusion of alternative BBN models 
is necessary in order to get more precise results and predictions than 
the ones presented here. In this paper we have carried out preliminary 
calculations that show
the viability of using HD molecule lines to search for signatures of early 
structure formation in the universe and/or to constrain cosmological theories.

\section*{ACKNOWLEDGMENTS}

We acknowledge the referee, Dr. N. Yoshida, for his accurate report and
useful comments and suggestions. We are grateful to J. Benda for grammar 
corrections to the manuscript. This work was supported by PROMEP grant 
12507 0703027 to A.L. and DGAPA-UNAM IN107706-3 grant to V.A.
A CONACyT PhD Fellowship is acknowledged by R.N-L.


\end{document}